\documentclass[a4paper,12pt,twoside=false]{article}

\usepackage[utf8]{inputenc} 
\usepackage{amsmath, amssymb} 
\usepackage{amssymb}
\usepackage{graphicx}
\usepackage{geometry} 
\usepackage{booktabs} 
\usepackage{enumitem} 
\usepackage{xcolor} 
\usepackage{hyperref} 
\usepackage{listings} 
\usepackage{mdframed} 
\usepackage{footmisc} 
\usepackage{multirow}
\usepackage{adjustbox}
\usepackage{tcolorbox}

\definecolor{codegray}{rgb}{0.5,0.5,0.5}
\definecolor{codepurple}{rgb}{0.58,0,0.82}
\definecolor{backcolour}{rgb}{0.95,0.95,0.92}
\definecolor{darkgreen}{rgb}{0.0, 0.5, 0.0}
\definecolor{codeblue}{rgb}{0.25,0.5,0.75}

\lstdefinestyle{mystyle}{
    backgroundcolor=\color{backcolour},
    basicstyle=\ttfamily\footnotesize,
    commentstyle=\color{codegray},
    keywordstyle=\color{codeblue},
    numberstyle=\tiny\color{codegray},
    stringstyle=\color{codepurple},
    frame=none, 
    rulecolor=\color{gray},
    breaklines=true, 
    keepspaces=true, 
    numbers=left, 
    numbersep=5pt, 
    showspaces=false, 
    showstringspaces=false, 
    showtabs=false, 
    tabsize=2, 
    xleftmargin=1.5em, 
    xrightmargin=1.5em 
}

\title{\textbf{A Framework for Building Enviromics Matrices in Mixed Models}}
\author{
    \begin{tabular}[t]{cc}
        Bruno A. Trevisan\textsuperscript{1} & Vinícius S. Junqueira\textsuperscript{2} \\ Bruna de M. Florêncio\textsuperscript{3}  & Alexandre S. G. Coelho\textsuperscript{4} \\ Gustavo E. Marcatti\textsuperscript{5} & Rafael T. Resende\textsuperscript{6}
    \end{tabular}
}
\date{January 07, 2025}

\begin{document}

\maketitle

\renewcommand{\thefootnote}{\arabic{footnote}} 
\setlength{\footnotemargin}{0pt} 
\footnotetext[1]{Master's student in Genetics and Plant Breeding, School of Agronomy, Federal University of Goiás (UFG), 74.690-900, Goiânia, Brazil.}
\footnotetext[2]{Researcher, Bayer Crop Science, 38407-049, Uberlândia, Minas Gerais, Brazil.}
\footnotetext[3]{Undergraduate student in Agronomy, School of Agronomy, UFG, 74.690-900, Goiânia, Brazil.}
\footnotetext[4]{Professor, School of Agronomy, UFG, 74.690-900, Goiânia, Brazil.}
\footnotetext[5]{Professor, Federal University of São João del Rei (UFSJ), Forest Engineering Department, 35701-970, Sete Lagoas, Minas Gerais, Brazil; Researcher at \href{https://www.linkedin.com/company/thecrop}{TheCROP}.}
\footnotetext[6]{Professor, School of Agronomy, UFG, 74.690-900, Goiânia, Brazil; Researcher at \href{https://www.linkedin.com/company/thecrop}{TheCROP}.\\ Corresponding author: \texttt{rafael.tassinari@ufg.br}}

\vspace{-0.75cm}\begin{mdframed}
\begin{footnotesize}\vspace{0.25cm}
 {\bf Abstract.} This study introduces a framework for constructing enviromics matrices in mixed models to integrate genetic and environmental data to enhance phenotypic predictions in plant breeding. Enviromics utilizes diverse data sources, such as climate and soil, to characterize genotype-by-environment (G$\times$E) interactions. The approach employs block-diagonal structures in the design matrix to incorporate random effects from genetic and envirotypic covariates across trials. The covariance structure is modeled using the Kronecker product of the genetic relationship matrix and an identity matrix representing envirotypic effects, capturing genetic and environmental variability. This dual representation enables more accurate crop performance predictions across environments, improving selection strategies in breeding programs. The framework is compatible with existing mixed model software, including rrBLUP and BGLR, and can be extended for more complex interactions. By combining genetic relationships and environmental influences, this approach offers a powerful tool for advancing G$\times$E studies and accelerating the development of improved crop varieties.\\ \vspace{-0.25cm}
 
 \end{footnotesize}\vspace{0.25cm}
\end{mdframed}

\section{Introduction}

The omics alphabet has grown rapidly with advancements in high-throughput typing technologies (i.e., tools capable of efficiently measuring and analyzing large-scale data with high precision), encompassing fields such as genomics, transcriptomics, proteomics, metabolomics, phenomics, and so on. Recently, enviromics has emerged as a valuable addition, by providing a framework for characterizing the ``envirome''—the full set of environmental factors affecting biological processes. Initially applied in fields like psychiatry (Anthony 1995)\cite{anthony1995}, enviromics can be expanded to areas such as agriculture or plant biology (Teixeira et al., 2011)\cite{teixeira2011}. Integrating diverse environmental data sources, including climate, soil properties, air quality, and socioeconomic factors, captures the micro- and macro-environmental influences on complex traits. This approach offers a more comprehensive understanding of how environmental components interact with biological systems, making it a powerful tool across disciplines.

\newpage
The genetic basis of producing a phenotype is complex and involves the activation and interaction of several genes and metabolic pathways. The most significant benefit of enviromics in the context of breeding is the possibility of modeling additional features beyond phenotypes and genetic markers when predicting future crop performance. In plant and animal breeding, the concept of enviromics was first introduced in a \href{https://doi.org/10.1101/726513}{2019 bioRxiv preprint} (Resende et al., 2021)\cite{resende2021}. Enviromics is particularly valuable for breeding studies, as it helps breeders and scientists assess the various factors that influence crop performance under different environmental conditions. In this context, ``environmental conditions'' encompasses more than just geographical factors, such as the specific plots, fields, and locations where crops are planted. Instead, enviromics considers any information derived from Earth surface, atmospheric, remote sensing, and socioeconomic factors that may contribute to changes in genotypic performance across multiple locations. Enviromics is especially valuable for studying genotype-by-environment-by-management (G$\times$E$\times$M) interactions, helping to decouple diverse envirotypic factors affecting crop performance (Costa-Neto \& Fritsche-Neto, 2021; Crossa et al., 2021; Resende et al., 2021)\cite{costa-neto2021a, crossa2021, resende2021}. This capability optimizes breeding strategies and decisions, enabling breeders and scientists to make informed site-specific recommendations.

Many advancements in modeling G$\times$E (and/or G$\times$E$\times$M) interactions highlight the benefits of integrating environmental matrix structures with genetic data to improve predictive accuracy in breeding. Crossa et al. (2006)\cite{crossa2006} initially focused on genetic covariances for G$\times$E modeling. Jarquín et al. (2014)\cite{jarquin2014} employed reaction norm models that combined high-dimensional genomic and environmental data, and Cuevas et al. (2017)\cite{cuevas2017} used Bayesian kernel methods to refine G$\times$E predictions by accounting for genotype-specific responses to environments. Hadamard and Kronecker products have been shown to model covariance structures in genotype-by-environment interactions effectively, capturing complex relationships in crop trials (Martini et al., 2020)\cite{martini2020}. Costa-Neto et al. (2021)\cite{costa-neto2021b} incorporated enviromic data into nonlinear kernel-based models, broadening the scope of environmental variability considered. These developments underscore the shift from purely genetic models to approaches that integrate comprehensive environmental descriptors, aligning with the principles of enviromics.

The use of enviromics in breeding is vast, as it can be applied to develop new cultivars within breeding programs. Still, it is also beneficial for activities that follow after the new cultivars are identified (Resende et al., 2024a)\cite{resende2024a}. For instance, identifying the most suitable environment for multiplying these cultivars is transformative, as it can enhance the profitability of seed production while minimizing seed multiplication costs. Production costs are a key factor influencing seed prices for customers, and since genetics vary in their adaptability to environments, multiplying seeds in less optimal conditions can increase costs (Gevartosky et al., 2023)\cite{gevartosky2023}. While our intention is not to exhaust all possible applications, we emphasize that enviromics is a powerful tool extending beyond performance prediction. Its benefits span from developing new cultivars to optimizing seed multiplication, defining agronomic recommendations, and guiding in-farm practices aimed at maximizing the yield potential of the cultivars.

This document outlines the process of building enviromic matrices for mixed models, aiming to integrate these tools into widely used software packages such as BGLR (Pérez \& de Los Campos, 2014)\cite{perez2014} and rrBLUP (Endelman, 2011)\cite{endelman2011}. The preparation of this material was inspired by the work of Bates et al. (2014)\cite{bates2014}, specifically adapted here to assist Enviromics users in applying linear algebra to handle random multiple regressions. Please, note that, a solid understanding of matrix algebra is required for accurate enviromic statistical modeling (see for instance Searle \& Khuri, 2017)\cite{searle2017}. These matrices are pivotal in mixed models that analyze G$\times$E interactions and enable the integration of enviromic data into genomic selection frameworks. Properly constructed matrices facilitate the modeling of intricate environmental and genetic relationships, improving the prediction accuracy of phenotypic performance and ultimately accelerating the development of improved plant varieties.

\section{Enviromics Modeling Documentation}

Enviromics provides a robust framework for modeling genotype-by-environment interactions by integrating multiple sources of variation. It allows breeders to account for complex environmental influences and genetic relationships simultaneously, enabling more accurate prediction of phenotypic responses under varying conditions. The approach uses a block-diagonal structure for the \( \mathbf{Z} \) matrix to include random effects associated with different genotypic and envirotypic covariates across trials, offering flexibility in capturing site-specific and genotype-specific responses. The covariance structure $\mathbf{\Sigma \otimes A}$, where $\mathbf{\Sigma}$ represents variance-covariance among environmental effects, and A denotes the additive relationship matrix. This dual representation allows us to predict crop performance across diverse environments, optimizing selection strategies in breeding programs.

\begin{itemize}[leftmargin=1.5cm]
    \item \textbf{Model Complexity}: The enviromics approach integrates multiple $Z_i$ submatrices (e.g., $Z_1, Z_2, \ldots, Z_n$) into a block-diagonal matrix $\mathbf{Z}$ (Bates et al., 2014)\cite{bates2014}, each corresponding to a different trial or environmental condition. This structure provides a flexible framework for modeling random effects across multiple genotypes, accommodating diverse environmental scenarios and trial conditions.

    \item \textbf{Block-Diagonal $\mathbf{Z}$}: The matrix $\mathbf{Z}$ combines various random effects, capturing the structure of the data in the model:
    \[
    y = Xb + Z u + e, \quad \mathbf{Z} = \begin{bmatrix}
    Z_1 & 0 & \cdots & 0 \\
    0 & Z_2 & \cdots & 0 \\
    \vdots & \vdots & \ddots & \vdots \\
    0 & 0 & \cdots & Z_n \\
    \end{bmatrix}
    \]
    where $y$ is the response vector, $X$ is the design matrix for fixed effects, $b$ represents the fixed effects, $Z$ is the design matrix for random effects, $u$ is the vector of random effects, and $e$ is the residual error. Each block $Z_i$ corresponds to the design matrix for a specific genotype and combination of environmental conditions.

    \item \textbf{Random Effects Vector}: The random effects vector $u$ encapsulates the responses of genotypes under different envirotypic covariates, where each $u_{i,j}$ represents the effect of genotype $i$ and envirotypic covariate $j$:
    \[
    u^\top = [u_{1,1}, u_{1,2}, \ldots, u_{n,m}], \quad \text{with} \quad u \sim N(0, A \otimes \Sigma).
    \]
    This formulation allows for modeling the joint distribution of genetic and environmental factors, enhancing the accuracy of predictions.

    \item \textbf{Covariance Structure}: The envirotypic covariance matrix $\Sigma$ can be represented as the identity matrix $I_{m+1}$, where $m$ is the number of envirotypic covariates plus one random intercept (of the genotype), assuming no additional variance within each environment. The matrix $\mathbf{K}$ is defined as the Kronecker product of the kinship matrix $A$ and the identity matrix $I_{m+1}$, expressed as $\mathbf{K} = A \otimes I_{m+1}$. This structure allows for modeling genetic relationships among genotypes while accounting for environmental effects, leading to a comprehensive framework for predicting phenotypic outcomes in mixed models.

\end{itemize}

\newpage
The matrix $\mathbf{Z}$ must have dimensions matching the data structure in mixed models. The number of rows of this matrix should match the dimension of the number of observations, and the number of columns should align with the number of parameters to estimate. Then, the number of columns must also align with the number of rows (or columns) of the kernel matrix (e.g., kinship or similarity matrix). Such alignment ensures proper matrix multiplication and integration of random effects into the model.

\subsection{Enviromics Prediction}
The enviromics prediction model aims to accurately estimate phenotypic responses by regressing random effects associated with genotypes and multiple envirotypic covariates. The general prediction equation for multiple genotypes can be expressed as:

\[
\hat{\mathbf{y}} = \mathbf{X}\hat{\beta} + \mathbf{Z}\hat{\mathbf{u}}
\]

where $\hat{\mathbf{y}}$ is the vector of predicted phenotypic responses for all genotypes; $\mathbf{X}$ is the design matrix for fixed effects; $\hat{\beta}$ is the vector of estimated fixed effects (i.e., BLUE in the frequentist paradigm); $\mathbf{Z}$ is the design matrix for random effects, with dimensions $n \times m$, where $n$ is the number of observations and $m$ is the number of random effects; and $\hat{\mathbf{u}}$ is the vector of estimated random effects (i.e., BLUP in frequentist paradigm), containing the random effects for all genotype-environment combinations. And the specific prediction for each genotype \(i\) can then be written as:

\[
\hat{y}_{i} = \beta_0 + \sum_{j=0}^{k} \hat{u}_{ij} \cdot \text{EC}_j
\]

where $\hat{y}_{i}$ is the predicted phenotypic response for genotype $i$; $\beta_0$ is the fixed intercept; $\hat{u}_{ij}$ represents the estimated random effect for the $j$-th envirotypic covariate (EC) and genotype $i$; $\text{EC}_j$ is the value of the $j$-th envirotypic covariate, with $\text{EC}_0 = 1$ to account for the random intercept; and $k$ is the total number of envirotypic covariates considered in the model.

The predicted phenotypic responses for each genotype \(i\) (where \(i\) varies from \(1\) to \(n\)) are summarized in Table \ref{tab:enviromics_prediction}. Each equation incorporates random effects associated with envirotypic covariates \(j\) (where \(j\) varies from \(1\) to \(m\)). This allows for a comprehensive understanding of how genetic and environmental factors contribute to phenotypic variation.

\begin{table}[h]
    \centering
    \begin{tabular}{cl}
        \hline
        \textbf{Genotype} & \multicolumn{1}{c}{\textbf{Equation}} \\
        \hline
        1 & $\hat{y}_{1} = \beta_0 + \hat{u}_{1,0} + \hat{u}_{1,1} \cdot \text{EC}_1 + \hat{u}_{1,2} \cdot \text{EC}_2 + \ldots + \hat{u}_{1,m} \cdot \text{EC}_m$ \\
        
        2 & $\hat{y}_{2} = \beta_0 + \hat{u}_{2,0} + \hat{u}_{2,1} \cdot \text{EC}_1 + \hat{u}_{2,2} \cdot \text{EC}_2 + \ldots + \hat{u}_{2,m} \cdot \text{EC}_m$ \\
        
        \vdots & \vdots \\
        
        $n$ & $\hat{y}_{n} = \beta_0 + \hat{u}_{n,0} + \hat{u}_{n,1} \cdot \text{EC}_1 + \hat{u}_{n,2} \cdot \text{EC}_2 + \ldots + \hat{u}_{n,m} \cdot \text{EC}_m$ \\
        \hline
    \end{tabular}
    \caption{Predicted phenotypic responses for genotypes incorporating random effects from envirotypic covariates.}
    \label{tab:enviromics_prediction}
\end{table}

Envirotypic covariates (ECs) are measurable environmental factors that influence phenotypic performance, while ``enviromic markers'' represent these covariates in predictive models, similar to genomic markers (Resende et al., 2024a)\cite{resende2024a}. In this didactic work, we will consider ECs as enviromic markers, acknowledging the risk of potentially missing important nonlinear effects. To address such effects, see Costa-Neto et al. (2021)\cite{costa-neto2021b}.

\section{Data Description (a Toy Example)}
To exemplify the enviromics modeling, we will use genotypic observations across various locations with associated envirotypic covariates (e.g., temperature, soil properties, and precipitation). This information forms the basis for building design matrices and covariance matrices in Enviromics. Each location represents a different trial, with varying numbers of genotypes. The plotted points correspond to the geographic coordinates (Longitude and Latitude), while the labels (G$_A$, G$_B$, G$_C$, or G$_D$) indicate the genotypes (`\texttt{gen}') available at each trial site.

The first step involves loading the dataset with observations across multiple locations and genotypes. The dataset includes envirotypic covariates (EC$_1$, EC$_2$, EC$_3$, EC$_4$, EC$_5$) that vary by location and may influence the response of different genotypes. Each row represents a unique combination of location and genotype, providing the necessary structure for building the design matrices and the covariance matrix. The column named `\texttt{y}' represents the phenotypic variable (trait), which can be associated with crop production or productivity (whatever). The following code snippet demonstrates how to read the dataset into the R environment and display its contents for initial inspection (see the R chunk below):

\begin{mdframed}[linecolor=gray, linewidth=0.5pt, roundcorner=5pt]
\begin{lstlisting}[language=R]
# Read the data from a file
dat <- read.table("dat.txt", header = TRUE)

# Display the data
print(dat)
    loc lon lat gen  EC1    EC2    EC3    EC4     EC5    y
1   L1   1   4   GA -0.230 -1.265  0.111 -0.050  -0.270  12.88
2   L5   2   5   GA  0.129  1.224  0.498  0.867   0.968  17.83
3   L6   3   6   GA  1.715  0.360 -1.967  1.061  -0.132  18.41
4   L2   2   2   GB -0.560  0.461  0.401  0.391   0.472  12.08
5   L1   1   4   GB -0.230 -1.265  0.111  0.022  -0.046  11.84
6   L4   4   4   GB  0.071 -0.446  1.787  0.580   0.710	 15.82
7   L6   3   6   GB  1.715  0.360 -1.967  0.832  -0.486	 18.16
8   L2   2   2   GC -0.560  0.461  0.401  0.588   0.322	 12.34
9   L3   3   3   GC  1.559 -0.687 -0.556  0.522   0.166	 16.11
10  L2   2   2   GD -0.560  0.461  0.401  0.770   0.692	 15.25
11  L1   1   4   GD -0.230 -1.265  0.111  0.108  -0.142	 14.91
12  L4   4   4   GD  0.071 -0.446  1.787  0.607   1.262	 19.24
13  L5   2   5   GD  0.129  1.224  0.498  0.683   0.706	 18.16
14  L6   3   6   GD  1.715  0.360 -1.967  1.052  -0.299	 16.18
\end{lstlisting}
\end{mdframed}

\vspace{0.5 cm}

\begin{figure}[h]
    \centering
    \includegraphics[width=0.9\textwidth]{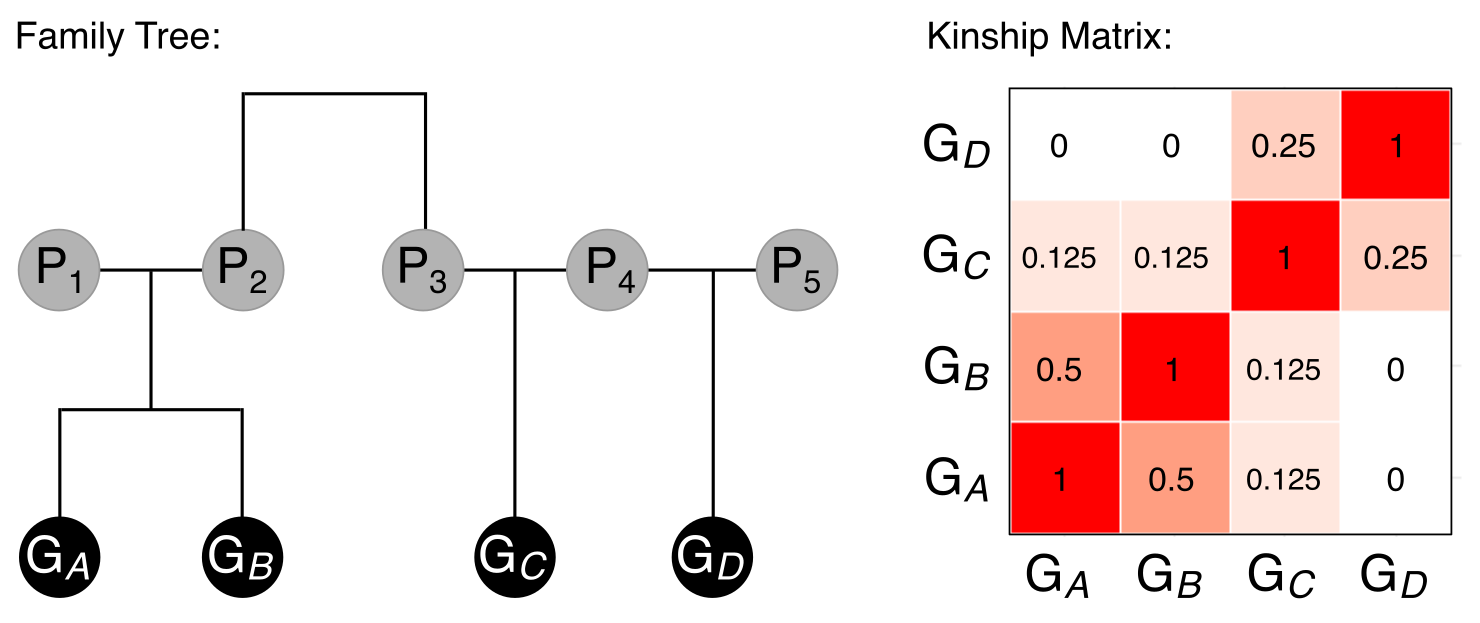}
    \caption{Pedigree and kinship matrix. The left side shows the genealogical relationships between the founders (P$_1$ to P$_5$) and their offspring (G$_A$, G$_B$, G$_C$, and G$_D$). The right side displays a heatmap of the kinship matrix, with colors representing the kinship coefficients between the genotypes.}
    \label{fig:ped_kinship}
\end{figure}

Figure~\ref{fig:ped_kinship} shows the pedigree of the four individuals and their relationship (kinship) matrix. On the left, the pedigree presents the relationships between founders (P$_1$ to P$_5$) and their offspring (genotypes G$_A$, G$_B$, G$_C$, and G$_D$). On the right, the relationship matrix quantifies the degree of relatedness between the genotypes, with colors representing the relatedness. The relationship matrix includes the coefficient of the relationship of Wright (1921)\cite{wright1921} and can be easily implemented following the Henderson rules to derive the inverse of this matrix directly (Henderson, 1976)\cite{henderson1976}.

\begin{tcolorbox}[colback=gray!5!white, colframe=gray!75!black, title={Note}]
\noindent The data used in this example consists solely of the above information. It is a small and didactic dataset that can be copied, pasted, and used directly.
\end{tcolorbox}

\newpage
\section{Preliminary Steps}
The initial steps for matrix preparation include the following:
\begin{itemize}
    \item The data is first sorted by genotype and then by location.
    \item Constructing block-diagonal design matrices to represent the structure of the observations.
    \item Build the relationship matrix based on specified genetic relationships, which will later be used to build kernel matrices.
\end{itemize}

\section{Enviromics Matrices Preparation}

Building enviromics matrices involves organizing data to capture genetic and environmental components in statistical models. The process begins with constructing a design matrix (\( Z \)) to represent random effects for each genotype and envirotypic covariate. This block-diagonal matrix accommodates multi-environment trials, allowing the model to separate effects across genotypes. A kernel matrix (\( K \)) also models genetic relationships (which is, in fact, merely an expanded kinship matrix), integrating genetic information with random effects through the Kronecker product for a comprehensive covariance structure. Together, these matrices form the foundation for mixed model analyses, enabling the prediction of phenotypic responses under various conditions. The process involves the following components:

\begin{itemize}
    \item \textbf{Design Matrix (\( Z \))}: A block-diagonal matrix where each block represents the design structure for a specific genotype, incorporating envirotypic covariates. The matrix captures the design structure for random effects with the following structure:
    \begin{mdframed}[linecolor=gray, linewidth=0.5pt, roundcorner=5pt]
    \begin{lstlisting}[language=R]
    # Split the data by genotype
    split_data <- split(dat, dat$gen)
    # Create the block-diagonal matrix Z
    Z <- bdiag(lapply(split_data, function(df) {
      # First column of 1's and columns for covariates:
      # EC1, EC2, EC3, EC4, EC5
      Z_i <- cbind(1, as.matrix(df[, c("EC1", "EC2", "EC3", "EC4", "EC5")]))
      Matrix(Z_i, sparse = TRUE)
    }))
    # Names the columns of matrix Z
    colnames(Z) <- unlist(lapply(1:4, function(i) {
        paste0("u", 0:5, "_G", LETTERS[i])
        }))
    # Display the matrix Z
    print(Z)
    \end{lstlisting}
    \end{mdframed}
\end{itemize}

Table~\ref{tab:design_matrix} shows the design matrix \( Z \) used in the mixed model. This matrix is constructed in a block-diagonal format, where each block represents the design effects for different genotypes (G$_A$, G$_B$, G$_C$, G$_D$). The first column of each block represents an intercept (\(u_0\)), while the subsequent columns represent envirotypic covariates (\( u_0, u_1, u_2, u_3, u_4, u_5 \)). The block-diagonal structure reflects the separation of effects for different genotypes, allowing efficient modeling of specific random effects.

\begin{table}[ht]
\begin{adjustbox}{max width=\textwidth}
\begin{tabular}{cccccccccccccccc}
\hline
\multicolumn{4}{c}{\textbf{G$_A$}} & \multicolumn{4}{c}{\textbf{G$_B$}} & \multicolumn{4}{c}{\textbf{G$_C$}} & \multicolumn{4}{c}{\textbf{G$_D$}} \\ \hline
\textbf{\(u_0\)} & \textbf{\(u_1\)} & ... & \textbf{\(u_m\)} & \textbf{\(u_0\)} & \textbf{\(u_1\)} & ... & \textbf{\(u_m\)} & \textbf{\(u_0\)} & \textbf{\(u_1\)} & ... & \textbf{\(u_m\)} & \textbf{\(u_0\)} & \textbf{\(u_1\)} & ... & \textbf{\(u_m\)} \\ \hline
1 & -0.230 & ... & -0.270 & 0 & 0 & ... & 0 & 0 & 0 & ... & 0 & 0 & 0 & ... & 0 \\
1 & 0.129 & ... & 0.968 & 0 & 0 & ... & 0 & 0 & 0 & ... & 0 & 0 & 0 & ... & 0 \\
1 & 1.715 & ... & -0.132 & 0 & 0 & ... & 0 & 0 & 0 & ... & 0 & 0 & 0 & ... & 0 \\
0 & 0 & ... & 0 & 1 & -0.560 & ... & 0.472 & 0 & 0 & ... & 0 & 0 & 0 & ... & 0 \\
0 & 0 & ... & 0 & 1 & -0.230 & ... & -0.046 & 0 & 0 & ... & 0 & 0 & 0 & ... & 0 \\
0 & 0 & ... & 0 & 1 & 0.071 & ... & 0.710 &  0 & 0 & ... & 0 & 0 & 0 & ... & 0 \\
0 & 0 & ... & 0 & 1 & 1.715 & ... & -0.486 & 0 & 0 & ... & 0 & 0 & 0 & ... & 0 \\
0 & 0 & ... & 0 & 0 & 0 & ... & 0 &  1 & -0.560 & ... & 0.322 & 0 & 0 & ... & 0 \\
0 & 0 & ... & 0 & 0 & 0 & ... & 0 &  1 & 1.559 & ... & 0.166 & 0 & 0 & ... & 0 \\
0 & 0 & ... & 0 & 0 & 0 & ... & 0 & 0 & 0 & ... & 0 & 1 & -0.560 & ... & 0.692 \\
0 & 0 & ... & 0 & 0 & 0 & ... & 0 & 0 & 0 & ... & 0 & 1 & -0.230 & ... & -0.142 \\
0 & 0 & ... & 0 & 0 & 0 & ... & 0 & 0 & 0 & ... & 0 & 1 & 0.071 & ... & 1.262 \\
0 & 0 & ... & 0 & 0 & 0 & ... & 0 & 0 & 0 & ... & 0 & 1 & 0.129 & ... & 0.706 \\
0 & 0 & ... & 0 & 0 & 0 & ... & 0 & 0 & 0 & ... & 0 & 1 & 1.715 & ... & -0.299 \\ \hline
\end{tabular}
\end{adjustbox}
\caption{Design matrix \( Z \) for the mixed model. Each block represents the design structure for a specific genotype (\(G_A\), \(G_B\), \(...\), \(G_D\)), with the first column (\(u_0\)) as an intercept, the last column (\(u_m\)) as a covariate, and ellipses (\(...\)) indicating omitted intermediate envirotypic covariates. In our case, \(m = 5\), corresponding to (\(u_0, u_1, ..., u_5\)).
}
\label{tab:design_matrix}
\end{table}

\begin{itemize}
    \item \textbf{Kernel Matrix (\( K \))}: Calculated using the Kronecker product, combining a genetic relationship matrix with an identity matrix representing different effects. This matrix is used to model the covariance structure between genotypes:
    \begin{mdframed}[linecolor=gray, linewidth=0.5pt, roundcorner=5pt]
    \begin{lstlisting}[language=R]
    # Create the relationship matrix based on pedigree
    A <- matrix(c(
      1.000, 0.500, 0.125, 0.000, 
      0.500, 1.000, 0.125, 0.000, 
      0.125, 0.125, 1.000, 0.250, 
      0.000, 0.000, 0.250, 1.000  
    ), nrow = 4, byrow = TRUE)

    # Define the genotype names
    rownames(A) <- colnames(A) <- unlist(lapply(1:4, function(i) {
    paste0("G", LETTERS[i])   }))

    # Create the 6x6 identity matrix for the effects (u0, u1, u2, u3, u4, u5)
    I6 <- diag(6)
    rownames(I6) <- colnames(I6) <- c("u0", "u1", "u2", "u3", "u4", "u5")

    # Calculate the Kernel matrix using the Kronecker product
    K <- kronecker(A, I6) # The resulting matrix will have dimensions 24x24

    # Rename the rows and columns of the K matrix to reflect genotypes and effects
    rownames(K) <- colnames(K) <- unlist(lapply(1:4, function(i) {
    paste0("u", 0:5, "_G", LETTERS[i])   }))
    \end{lstlisting}
    \end{mdframed}
\end{itemize}

\begin{figure}[ht]
    \centering
    \includegraphics[width=0.9\textwidth]{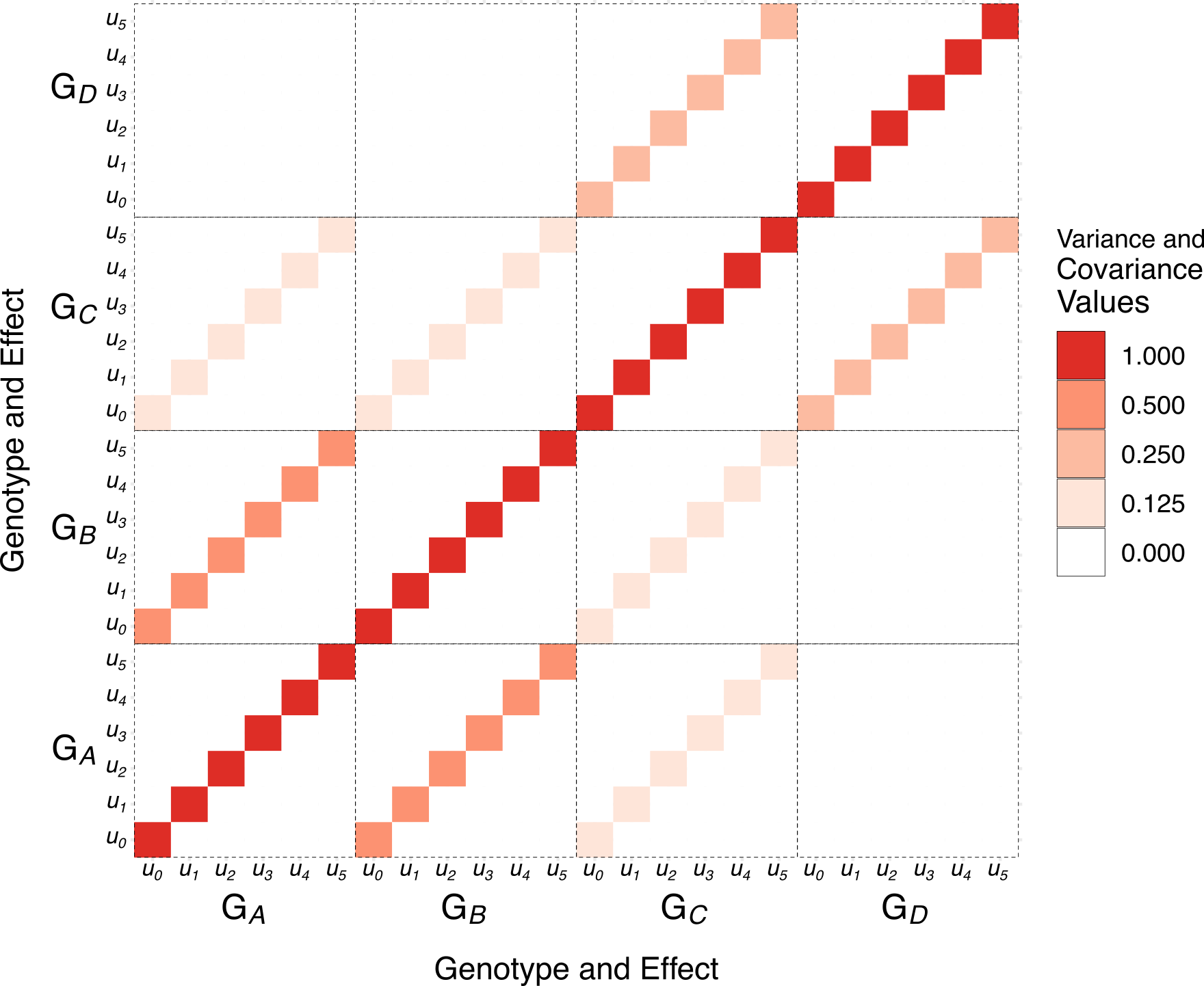}
    \caption{Heatmap visualization of the expanded kinship matrix \( K \), here also called as ``Kernel'' matrix, representing the genetic relationships across genotypes and effects. The matrix was generated using the Kronecker product to expand the genetic relationship matrix by different random effects (\( u_0, u_1, u_2, u_3, u_4, u_5 \)). This matrix is defined as the Kronecker product of the kinship matrix $A$ and the a identity matrix $I$.}
    \label{fig:expanded_A_matrix}
\end{figure}

\newpage
Figure~\ref{fig:expanded_A_matrix} illustrates the expanded kinship matrix \( K \), which combines the genetic relationship matrix with the Identity ($I_{m+1}$) matrix for different effects. The matrix \( K \) is constructed using the Kronecker product, capturing the covariance structure between genotypes (G$_A$, G$_B$, G$_C$, and G$_D$) across various random effects (\( u_0, u_1, u_2, u_3, u_4, u_5 \)). The heatmap visualizes the kinship values, with a color gradient indicating the degree of relatedness.

In the context of the kernel matrix, it is assumed that there is no relationship between the random effects \( u_0, u_1, u_2, u_3, u_4, u_5 \) within each genotype. This follows Fisher's infinitesimal model, which posits that numerous small, independent effects contribute to the overall genetic variation. In practical applications of enviromics, the model can accommodate thousands of such effects (\( u_1, u_2, \ldots, u_m \)), representing various envirotypic covariates. However, when considering specific effects across genotypes, such as \( u_0 \times u_0 \), \( u_1 \times u_1 \), \( u_2 \times u_2 \), \( u_3 \times u_3 \), \( u_4 \times u_4 \), and \( u_5 \times u_5 \), there is a structured relationship that reflects the genetic relatedness specified by the kinship matrix. This kinship can be based on pedigree information or derived from molecular markers like SNPs (VanRaden, 2008)\cite{vanraden2008}, allowing the model to incorporate genetic relationships across different random effects.

The matrix methodology presented here is similar to that described in Resende et al. (2024b) \cite{resende2024b}. This model was applied using the \texttt{rrBLUP} package (Endelman, 2011)\cite{endelman2011} for a frequentist approach and the \texttt{BGLR} package (Pérez \& de Los Campos, 2014)\cite{perez2014} for a Bayesian approach, as both allow flexible manipulation of matrices. However, many other R packages can be used for this purpose, depending on the user's preferences and familiarity. To understand the predictive performances of the models described here, please refer to the studies by Resende et al., (2021)\cite{resende2021}, and Resende et al., (2024b)\cite{resende2024b}.

\section{Enviromics Matrices: Frequentist and Bayesian Approaches}

The study of genetic and non-genetic components of populations can be approached through a variety of methodologies. From a statistical perspective, these methods are broadly classified into supervised and unsupervised approaches. Supervised methods, in particular, can be further divided into parametric and non-parametric categories, each with distinct assumptions and analytical frameworks. In this work, we focus on supervised parametric approaches within the frequentist and Bayesian paradigms. These methods have been extensively applied over decades to investigate the genetic architecture of plants, animals, microorganisms, and human populations, contributing to significant advancements in quantitative genetics and breeding. In the following sections, we delve into the technical foundations of these parametric approaches, with an emphasis on genotype-by-environment interaction modeling. Additionally, we illustrate their application through a simple example, demonstrating the construction of design matrices in the emerging field of enviromics.

For a large number of envirotypic covariates (ECs), convergence issues may arise both in Frequentist and Bayesian approaches. To address this, `ensemble models' can be applied by grouping ECs and running separate models for each group in enviromics models (Resende et al., 2024b)\cite{resende2024b}. Going further, in practical / empirical applications, it is also essential to cross-validate these models by implementing training and validation sets. For this purpose, we recommend exploring strategies for multi-environment and/or multi-year validations across regions, as discussed in Rogers \& Holland (2022)\cite{rogers2022}.

\newpage
\subsection{Frequentist approach}

Using supervised parametric methods in genetics study involves estimating parameters that describe the source variations of a given population. Over the years, several techniques, such as Methods I, II, and III of Henderson (1953)\cite{henderson1953}, Methods MINQUE and MIVQUE of Rao (1970, 1971a, 1971b)\cite{rao1970,rao1971a,rao1971b}, ANOVA, maximum likelihood (ML), and restricted (or residual) maximum likelihood (REML) (Patterson \& Thompson, 1971)\cite{patterson1971}, were developed to help disentangle the source of variations into different components. Each one of these methods aims to optimize quadratic forms under different assumptions related to the data structure (i.e., balanced and unbalanced), as well as the bias and variance of estimates. Currently, the most widely used method is REML due to its reliability and flexibility in managing complex data structures and its ability to produce unbiased estimates for fixed and random effects.

As long as the variance components are available, estimating the genetic values becomes a (relatively) trivial task under Henderson's mixed model equations. However, in practice, the estimation of variance components and genetic value prediction occurs in a two-step iterative process. The first step estimates the variance components through maximum likelihood or restricted maximum likelihood methods. Then, the fixed and random effects are derived in the second step, assuming that the variance components estimated in the first step are true. The model converges to a stable solution after a few rounds of iterative process. Currently, several computationally efficient algorithms can execute this task.

In the Genomic Era, mixed models have been used successfully to predict genomic values in many species. Nowadays, multiple models are used, and they can differ if the goal is to estimate the genetic values or marker (e.g., SNP) effects. Among the methods, the ridge regression BLUP (RRBLUP) is a popular one. RRBLUP employs ridge regression to handle multicollinearity issues common in high-dimensional datasets, such as those encountered in genomics, where usually there are way more markers than observations. RRBLUP works by partitioning phenotypic variation into genetic and non-genetic components using mixed models, where fixed effects account for general trends and random effects capture genotype-specific variations. Fixed effects capture consistent influences shared across all genotypes, while random effects model genotype-specific deviations from the population mean. This framework integrates a design matrix (\(Z\)), a kernel matrix (\(K\)), and phenotypic values (\(y\)) to effectively model genotype-by-environment interactions. The estimated fixed and random effects are organized into structured data frames, providing a systematic approach for downstream analysis and interpretation (Burgueño et al., 2007 \cite{burgueno2007}).

The following code demonstrates the application of the \texttt{rrBLUP} R package to partition phenotypic variation into genetic and environmental components. The phenotypic data (\(y\)) is stored, and the model is fitted using the design matrix (\(Z\)) and the kernel matrix (\(K\)). Fixed effects are extracted and stored for further interpretation, while random effects are arranged and labeled in a structured data frame for downstream analysis.

    \begin{mdframed}[linecolor=gray, linewidth=0.5pt, roundcorner=5pt]
    \begin{lstlisting}[language=R]
    # Storing phenotypic data
    y <- as.matrix(dat$y)
    
    # Fitting the rrBLUP model 
    rrblupmodel <-  mixed.solve (y = y, Z = Z, K = K)

    # Storing fixed effect
    fixef_rrblup <- as.data.frame(rrblupmodel$beta)[, 1]

    #Storing, arranging and naming random effects
    ranef_rrblup <- as.data.frame(rrblupmodel$u)[, 1]
    ranef_matrix_rrblup <- matrix(ranef_rrblup, nrow = 4, ncol = 6, byrow = TRUE)
    ranef_df_rrblup <- as.data.frame(ranef_matrix_rrblup)
    rownames(ranef_df_rrblup) <- c("GA","GB","GC","GD")
    colnames(ranef_df_rrblup) <- c("u_0","u_1","u_2",
                                   "u_3","u_4","u_5")
    \end{lstlisting}
    \end{mdframed} 

The observed effects shown in Figure~\ref{fig:predictions} were predicted using the mixed model equation \(\hat{\mathbf{y}} = \mathbf{X}\hat{\beta} + \mathbf{Z}\hat{\mathbf{u}}\), where \(\mathbf{X}\hat{\beta}\) represents the fixed effects and \(\mathbf{Z}\hat{\mathbf{u}}\) accounts for the random effects. These components were extracted and arranged using the \texttt{rrBLUP} package, as shown in the provided code. The fixed and random effects for each genotype and envirotypic index were then combined to generate the predicted values plotted in the figure.

Using the fitted models, phenotypic responses are predicted for each genotype across a gradient of envirotypic indices. These predictions combine fixed effects, representing baseline responses, with random effects, accounting for genotype-specific deviations. This dual-component structure enables a nuanced understanding of how genotypes respond to varying environmental conditions, as in Table~\ref{tab:enviromics_prediction}. Such modeling captures the complexities of genotype-by-environment interactions, facilitating the identification of genotypes with optimal performance under specific conditions (Cuevas et al., 2018)\cite{cuevas2018}.

\subsection{Bayesian approach}

Bayesian framework is another method commonly used to estimate variance components and derive genetic values. Bayesian methods date back to Thomas Bayes, who was credited with developing Bayes' Theorem, a fundamental probability theory and statistics principle. However, the broader Bayesian paradigm, as we know it today, was shaped and expanded by many contributors over time. Such continued contributions by other scientists enabled the utilization of the Bayesian approach in practical conditions. The central principle of the Bayesian approach is that the posterior probability of the parameters of interest—such as fixed and random effects and variance components—is determined by the joint distribution of the data, which is represented by the likelihood function alongside additional terms related to the prior probability of the model's parameters. The prior probability in the Bayesian framework reflects our a priori knowledge about a specific parameter. In the context of genomics, this prior information typically relates to the probability function that describes the distribution of genetic markers. For instance, if we assume that several genes of small effect influence a particular trait, a normal distribution may be a suitable model for that trait. Conversely, if a few genes of large effect control the trait under investigation, an exponential or t-distribution may better capture its characteristics than a normal distribution. Additionally, there is the option to adjust the functions that model the likelihood of a specific marker's inclusion—or exclusion—in influencing phenotypic expression. These models are referred to as Bayesian variable selection methods.

\begin{figure}[ht]
    \centering
    \includegraphics[width=\textwidth]{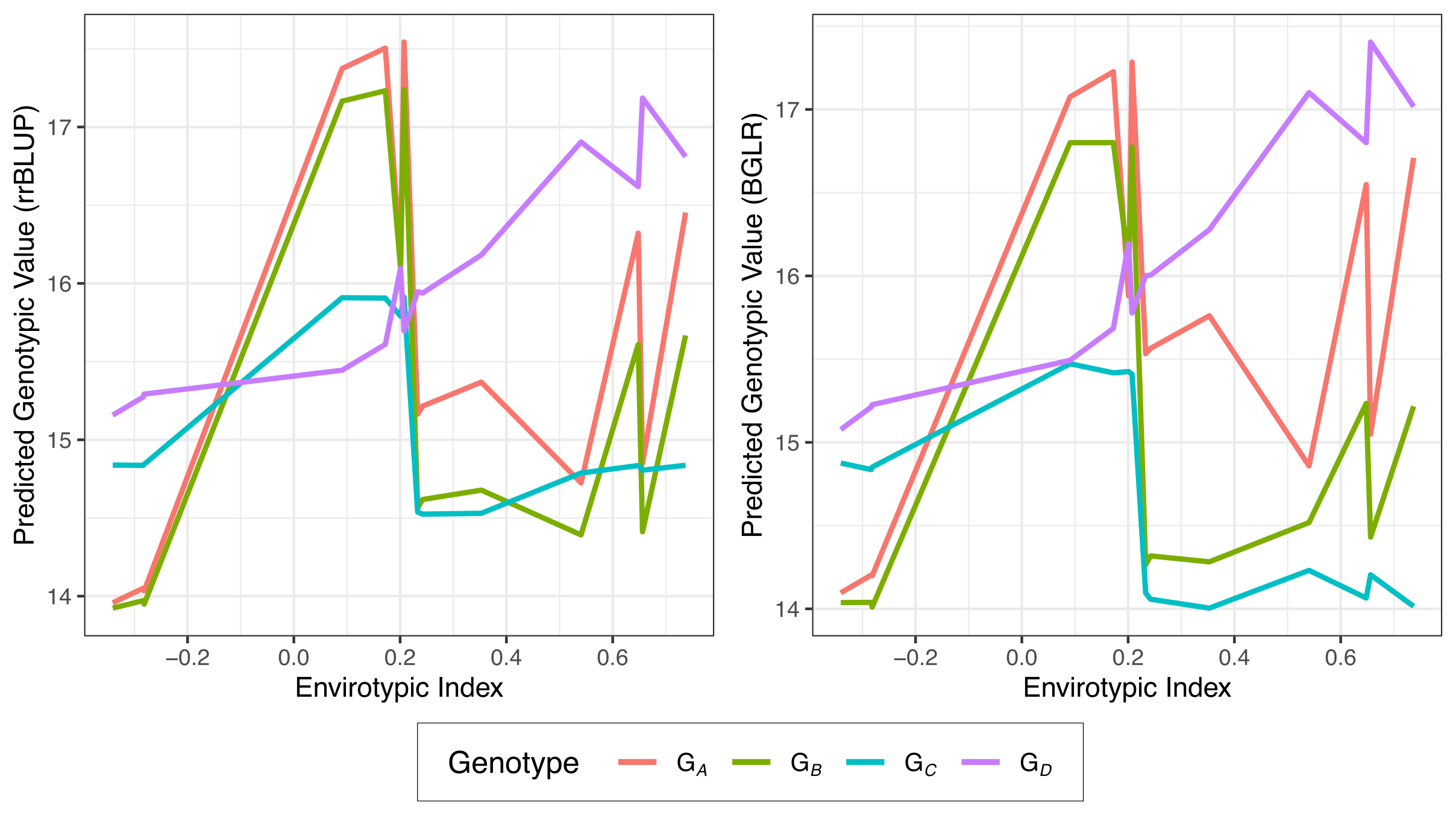}
    \caption{Comparison of Predicted Genotypic Values from the rrBLUP and BGLR models across an Envirotypic Index for different genotypes (G$_A$, G$_B$, G$_C$, and G$_D$). The ``Envirotypic Index'' was assumed to be simply the average of the five envirotypic covariates (EC$_1$--EC$_5$).}
    \label{fig:predictions}
\end{figure}

A key concept in the Bayesian approach is that all effects are treated as random. Each effect included in the model is represented using a specific probability distribution function that describes its behavior. In Bayesian analysis, when we want to indicate that a frequentist fixed effect is included explicitly, we refer to it as a systematic effect. Typically, these systematic effects are incorporated into the model using either a uniform or normal distribution with a very large variance. This approach allows us to assume that the effect remains relatively consistent across all levels of the other parameters in the model. In genetics, we need to explicitly define the probability function that will be used to describe either marker effects or genetic effects. The Bayesian framework is particularly advantageous for exploring the various hypotheses related to the genetic control of traits and their impact on predictions. There are many options available within this framework. One common option is to use a normal prior with equal variance for all markers, which is implemented in the RRBLUP method. In the Bayes A method, each marker is assigned its normal prior with a specific variance. Bayes B is similar to Bayes A but requires a predefined proportion ($\pi$) of markers to have non-zero effects, facilitating variable selection. Methods like Bayes C$\pi$ and D$\pi$ incorporate a probability function related to the proportion of markers with non-zero effects (Gianola, 2013)\cite{gianola2013}. These methods are improved versions of Bayes B, allowing the model to determine the optimal $\pi$ based on available data (likelihood function) and the assumed prior distributions. This aspect of Bayesian analysis is particularly beneficial as it enables a learning process during the evaluation.

Several computational packages are available for analyzing genomic data using Bayesian methods, such as \texttt{bWGR} (Xavier et al., 2020)\cite{xavier2020}. In this demonstration, we illustrate the application of Bayesian Ridge Regression (BRR) using the \texttt{BGLR} package. This versatile package allows users to choose from various probabilistic models and employs Markov Chain Monte Carlo (MCMC) methods via the Gibbs sampling algorithm for parameter estimation. In our model, we utilize a normal prior for the kernel matrix \(K\) and specify both the kernel matrix and the design matrix \(Z\) within the \texttt{ETA} list. The analysis is executed over 50,000 iterations, with a burn-in period set at 10\% of the total iterations. Systematic effects, treated as fixed effects in frequentist approaches, are estimated and recorded, while the remaining random effects are structured and labeled in a data frame for downstream analysis. This structured output is particularly useful for interpreting genotype-by-environment interactions. The BGLR framework enhances genomic analysis through its flexibility and ability to incorporate prior knowledge. Unlike deterministic estimators like rrBLUP, BGLR uses iterative sampling to estimate parameters, providing a quantification of uncertainty in predictions (Montesinos-López et al., 2016)\cite{montesinoslopez2016}. Our implementation adopts a normal prior distribution, consistent with frequentist assumptions, to enable direct comparisons between Bayesian and frequentist results. Following the completion of the MCMC process, the first step is to discard the burn-in samples to mitigate the influence of initial conditions. To further address autocorrelation often present in MCMC samples, we retain only one sample for every n-th iteration (thinning). After this initial processing, it is critical to evaluate the convergence of the retained samples. Convergence diagnostics typically involve graphical inspections (e.g., trace plots) and statistical tests, such as the Geweke diagnostic, to ensure the reliability of parameter estimates.

    \begin{mdframed}[linecolor=gray, linewidth=0.5pt, roundcorner=5pt]
    \begin{lstlisting}[language=R]
    # Fitting the BGLR model 
    ETA <- list(A = list(X = Z, K = K, model = 'BRR'))

    niter <- 50000

    BGLRmodel <- BGLR(y = y, 
            ETA=ETA, 
            nIter = niter, 
            burnIn = niter*.10, 
            thin = 20, 
            saveAt = 'BGLRoutput')


    # Storing fixed effect
    fixef_bglr <- as.data.frame(BGLRmodel$mu)[, 1]

    #Storing, arranging and naming random effects
    ranef_bglr <- as.data.frame(BGLRmodel$ETA$A$b)[, 1]
    ranef_matrix_bglr <- matrix(ranef_bglr, nrow = 4, ncol = 6, byrow = TRUE)
    ranef_df_bglr <- as.data.frame(ranef_matrix_bglr)
    rownames(ranef_df_bglr) <- c("GA","GB","GC","GD")
    colnames(ranef_df_bglr) <- c("u_0","u_1","u_2","u_3","u_4","u_5")
    \end{lstlisting}
    \end{mdframed} 

Model performance is assessed by comparing predictions from rrBLUP and BGLR side-by-side in Figure~\ref{fig:predictions}. The high correlation (0.973) between predictions highlights their consistency in estimating genotypic values. This agreement underscores the robustness of these methods in capturing additive and additive-by-additive genetic interactions across environments (Solaymani et al., 2020)\cite{solaymani2020}. However, the Bayesian approach's ability to incorporate prior knowledge and quantify uncertainty offers additional flexibility, particularly for complex multi-environment trials involving multiple traits.

Bayesian analysis allows the integration of different sources of genetic information, such as the relationship matrix (\(A\)) and the genomic relationship matrix (\(G\)) for genomic-enviromic predictions (Cuevas et al., 2017)\cite{cuevas2017}. The possibility of using different priors can be a key feature that enhances the model's ability to derive reliable predictions. Informative priors are valuable in real-world breeding studies when prior knowledge exists about genetic relationships, environmental effects, or trait heritability. For instance, priors based on pedigree information can guide the model when genomic data is sparse, while priors derived from historical yield data can improve predictions in environments with limited phenotypic observations. Additionally, shrinkage priors, such as those used in Bayesian Ridge Regression, are effective for controlling overfitting when working with a large number of predictors. By incorporating these priors, \texttt{BGLR} allows breeders and scientists to leverage prior knowledge effectively, resulting in more accurate and reliable predictions for complex breeding scenarios.

\newpage
\section{Next Steps}
The next phase of implementing enviromic models involve:
\begin{itemize}
    \item Validating the matrix construction with real datasets.
    \item Extending the approach to account for more complex genetic and environmental interactions.
    \item Prediction of parents, for instance, when dealing with lines, to obtain specific hybrids.
    \item Interpolate the genotypic prediction across an entire framed area based on the ``Target Population of Environments'' (TPE) (Cruz et al., 2024)\cite{cruz2024} for the crop.
    \item Evaluating the impact of different covariance structures on the predictive performance of the mixed models.
\end{itemize}

\section{Final Considerations}

Developing enviromics matrices for mixed models using frequentist or Bayesian methods is a robust framework for integrating genetic and environmental data in plant breeding. The construction of design and kernel matrices enables the accurate modeling of genotype-by-environment interactions, allowing for improved prediction of phenotypic performance under diverse conditions. While the current approach uses a block-diagonal structure to separate random effects across genotypes and assumes no relationships between different \( u \)'s within the same genotype, future advancements could extend these methods to more complex scenarios.

Mathematical developments in matrix expansions could involve incorporating multi-trait models, where multiple phenotypic traits are analyzed simultaneously, accounting for correlations among them. Another promising direction is including diverse experimental types within a unified framework, allowing for the simultaneous analysis of trials with different designs, such as replicated trials, unreplicated trials, or even observational data. Additionally, introducing covariances between random effects \( u \)'s across genotypes could further enhance the modeling of shared environmental influences or interactions among traits.

Such expansions would require sophisticated matrix derivations and computational techniques, potentially involving sparse matrix algebra operations and high-dimensional statistical methods. By pursuing these advancements, enviromics could provide even more comprehensive insights into genetic and environmental interactions, driving more effective breeding strategies and accelerating the development of improved crop varieties. In addition, future research should focus on developing more computationally efficient strategies to enable large-scale analyses. As environmental data becomes widely available and included in the genetic evaluation of breeding programs, new software will be required to manage the amount of data, especially when integrated with additional omics information such as genomic, transcriptomic, proteomics, metabolomics, epigenomics.

\paragraph{Acknowledgments.} We express our gratitude to the PPGGMP/UFG (``Programa de Genética e Melhoramento de Plantas'' at UFG--``Universidade Federal de Goiás'') and the LAMP (``Laboratório de Melhoramento de Precisão'') for their support during the development of this work. We declare no conflicts of interest.

\vspace{0.5cm}\begin{mdframed}\vspace{-0.5cm}
\paragraph{Cite this article as:} Trevisan, B. A., Junqueira, V. S., Florêncio, B. de M., Coelho, A. S. G., Marcatti, G. E., \& Resende, R. T. (2024). {\bf A Framework for Building Enviromics Matrices in Mixed Models}. \textit{arXiv}.
\end{mdframed}

\end{document}